\documentclass[prd,preprintnumbers,twocolumn,eqsecnum,floatfix,nofootinbib,a4paper,superscriptaddress]{revtex4-1}
\usepackage{color}
\usepackage{calc}
\usepackage{amsmath,amssymb,graphicx}
\usepackage{amssymb,amsmath}
\usepackage{tensor}
\usepackage{bm}
\usepackage{float}
\usepackage{microtype}
\usepackage{booktabs}
\usepackage{times}
\usepackage[varg]{txfonts}
\usepackage[colorlinks, pdfborder={0 0 0}]{hyperref}
\usepackage{subfigure}
\definecolor{LinkColor}{rgb}{0.75, 0, 0}
\definecolor{CiteColor}{rgb}{0, 0.5, 0.5}
\definecolor{UrlColor}{rgb}{0, 0, 0.75}
\hypersetup{linkcolor=LinkColor}
\hypersetup{citecolor=CiteColor}
\hypersetup{urlcolor=UrlColor}
\allowdisplaybreaks
\usepackage[utf8]{inputenc}
\usepackage{ulem}
\usepackage{comment}
\usepackage{hyperref}
\usepackage[nolist]{acronym}
\usepackage{tikz}
\usepackage{textcomp}
\usepackage[export]{adjustbox}
\usepackage{url}

\usetikzlibrary{arrows,shapes,trees,decorations.pathreplacing}
\usepackage{aas_macros}

\allowdisplaybreaks

\begin{document}

\title{Prospects for probing ultralight primordial black holes using the stochastic gravitational-wave background induced by primordial curvature perturbations}

\author{Shasvath J. Kapadia}
\affiliation{International Centre for Theoretical Sciences, Tata Institute of Fundamental Research, Bangalore - 560089, India}

\author{Kanhaiya Lal Pandey}
\affiliation{International Centre for Theoretical Sciences, Tata Institute of Fundamental Research, Bangalore - 560089, India}

\author{Teruaki Suyama}
\affiliation{Department of Physics, Tokyo Institute of Technology, 2-12-1 Ookayama, Meguro-ku, Tokyo 152-8551, Japan}

\author{Parameswaran Ajith}
\affiliation{International Centre for Theoretical Sciences, Tata Institute of Fundamental Research, Bangalore - 560089, India}
\affiliation {Canadian Institute for Advanced Research,CIFAR Azrieli Global Scholar, MaRS Centre, West Tower, 661 University Ave, Toronto, ON M5G 1M1, Canada}


\begin{abstract}
Ultralight primordial black holes (PBHs) with masses $\lesssim 10^{15}$g and subatomic Schwarzschild radii, produced in the early Universe, are expected to have evaporated by the current cosmic age due to Hawking radiation. Based on this assumption, a number of constraints on the abundance of ultralight PBHs have been made. However, Hawking radiation has thus far not been verified experimentally. It would, therefore, be of interest if constraints on ultralight PBHs could be placed independent of the assumption of Hawking-radiation. In this paper, we explore the possibility of probing these PBHs, within a narrow mass range, using gravitational-wave (GW) data from the two LIGO detectors. The idea is that large primordial curvature perturbations that result in the formation of PBHs, would also generate GWs through non-linear mode couplings. These induced GWs would produce a stochastic background. Specifically, we focus our attention on PBHs of mass range $\sim 10^{13} - 10^{15}$g for which the induced stochastic GW background peak falls in the sensitivity band of LIGO. We find that, for both narrow and broad Gaussian PBH mass distributions, the corresponding GW background would be detectable using presently available LIGO data, provided we neglect the existing constraints on the abundance of PBHs, which are based on Hawking radiation. Furthermore, we find that these stochastic backgrounds would be detectable in LIGO's third observing run, even after considering the existing constraints on PBH abundance. A non-detection should enable us to constrain the amplitude of primordial curvature perturbations as well as the abundance of ultralight PBHs. We estimate that by the end of the third observing run, assuming non-detection, we should be able to place constraints that are orders of magnitude better than currently existing ones.
\end{abstract}

\maketitle

\section{Introduction}\label{sec:introduction}
Primordial black holes (PBHs) are thought to be produced via the direct collapse of overdense regions in the early Universe, in contrast to astrophysical black holes, which are produced by the collapse of the cores of massive stars. The lower limit on the allowed range of masses for PBHs is much smaller than for astrophysical black holes, permitting even the possibility of Planck mass ($\sim 10^{-5}$g) PBHs \cite{1971MNRAS.152...75H,1974MNRAS.168..399C}. 

PBHs have the potential to provide answers to many open questions in astrophysics and cosmology. They are among the often-considered candidates for dark matter, could aid the formation of supermassive black holes and galaxies, and are speculated to be possible sources of gamma-ray bursts \cite[and references therein]{2005tsra.conf...89C,2015PhRvD..92b3524C,2016PhRvD..94h3504C,2018JApA...39....9P}. If they are found to exist, they could also probe the primordial spectrum of density fluctuations and various phase transitions in the early Universe \cite{2005tsra.conf...89C,2018CQGra..35f3001S}.

They have therefore been searched for extensively using various observations. These include observations of the extra-galactic $\gamma$-ray background, gravitational microlensing experiments (e.g.,. OGLE-I-IV, using Kepler objects, Eridanus-II star clusters), cosmic microwave background experiments, dynamical constraints, and accretion constraints \cite[and the references therein]{2016PhRvD..94h3504C,2018CQGra..35f3001S,Carr:2020gox}. 
These observations and experiments provide strong constraints on PBH abundances in various mass windows above$\sim 10^{15}$g.
Those with masses $\lesssim 10^{15}$g, which we refer to here as ``ultralight PBHs'', are thought to have evaporated by now due to Hawking-radiation. 
For such PBHs  we have constraints coming from measurements of the abundances of light elements, since the presence of such PBHs 
in the primordial universe would change the abundance 
of light elements predicted in the standard big-bang nucleosynthesis
 (BBN) due to high energy particles of the Hawking radiation.
\cite[and the references therein]{2010PhRvD..81j4019C,2018CQGra..35f3001S}. 

A more recent means of probing PBHs is via gravitational waves (GWs). Ever since their detection on September 14, 2015, GWs have afforded a novel way to infer the existence of binary black holes (BBHs), as well as their intrinsic parameters such as their masses and spin angular momenta. During the first and second observing runs (O1 and O2) of the LIGO-Virgo detectors, there were more than a dozen confirmed BBH detections \cite{GWTC-1, Venumadhav2019}, most of which were significantly heavier than those observed in X-ray binaries. Their unexpectedly large masses provided ample opportunity for a number of theoretical formation channels to be proposed \cite{2016Natur.534..512B}. Among them is the possibility that (at least some) of LIGO's binary black holes are PBHs \cite{2015PhRvD..92b3524C,2016PhRvL.116t1301B,2016PhRvL.117f1101S} \footnote{It has also been speculated that the putative binary neutron star merger event GW190425  \cite{GW190425-Detection} could be a coalescence of two PBHs}. 

While ascertaining the provenance of LIGO's black holes as PBHs might be challenging, detecting mergers of sub-solar BBHs would likely prove to be an important step in establishing the existence of non-evaporated PBHs, since astrophysical black holes are not expected to be lighter than $\sim 3M_{\odot}$. The non-detection of sub-solar BBHs would constrain the fraction of dark-matter ($f_{\mathrm{PBH}}$) as non-evaporated sub-solar PBHs. A search for a narrow mass-range of resolvable sub-solar BBH systems in LIGO-Virgo data was recently conducted, and a corresponding constraint on $f_{\mathrm{PBH}}$ placed \cite{O1SubSolar, O1SubSolarMethods, 2018CQGra..35f3001S}. In addition, a search in O1 data for the stochastic background of GWs from coalescing stellar mass BBHs \cite{O1Stochastic} and sub-solar-mass BBHs \cite{Wangetal2018} was also carried out.

In this article, we explore the prospects of detecting a stochastic background of GWs from primordial curvature perturbations, which also result in the formation of PBHs through gravitational collapse. Owing to this close connection, these induced GWs offer an interesting probe of the abundance and mass function of PBHs~\cite{Saito:2008jc, Saito:2009jt}. Unlike most previous searches for PBHs via GWs (resolved or stochastic), we do not rely on these PBHs forming binaries. Instead, we focus on PBH formation in the radiation dominated era due to a large peak in the primordial spectrum of curvature fluctuations around the mass scale of $\sim 10^{13-15}$g. Such scenarios are possible in some hybrid inflation theories \cite[and the references within]{1996PhRvD..54.6040G,2015PhRvD..92b3524C,2018CQGra..35f3001S}. These curvature fluctuations would lead to the formation of GWs via scalar-tensor mode coupling in the second order perturbation theory, thus producing a stochastic GW background \cite[and the references therein]{2018CQGra..35f3001S, Wangetal2019}. We investigate the detectability of this stochastic background by the LIGO observatories with sensitivities achieved during the first three observing runs as well as the expected design sensitivity. 

We find that the GW energy density fraction $\Omega_{\mathrm{GW}}$ resulting from a narrow Gaussian peak in the curvature power spectrum should be detectable in O1 and O2 data, if we allow the primordial power spectrum to have its maximum possible amplitude (neglecting the existing constraints that are derived assuming Hawking radiation). However, a significant fraction of its signal-to-noise ratio (SNR) comes from a feature associated with narrow peaks (see sections \ref{sec:stochastic} and \ref{sec:perturbations} for details). While this might appear unrealistic in the context of single-field inflation models \cite{Byrnes:2018txb, Carrilho2019}, we nevertheless investigate its detectability,  
since certain multiple field models of inflation predict such narrow distributions \cite{1996PhRvD..54.6040G, Kawasaki:2006zv, Kawaguchi:2007fz, Kawasaki:2012kn, Palma:2020ejf}. 
On the other hand, $\Omega_{\mathrm{GW}}$ resulting from a broad Gaussian peak in the curvature power spectrum, is also detectable in O1 and O2 data provided Hawking radiation is ignored. 
%
Both spectra should be detectable in O3 data and data at Design sensitivity, even when existing Hawking-radiation based constraints are considered. These results advocate a search for the ultralight PBHs investigated in this paper in LIGO-Virgo data from O1 and O2, as well as O3. 

It is worth mentioning here that the amplitude of the energy-density fraction $\Omega_{\mathrm{GW}}$ (which comes from the curvature power spectrum) is a free parameter that needs to be constrained from observation. We consider two cases where we investigate detectability. The first uses existing constraints on the abundance of ultralight PBHs from BBN and extra-galactic photon background, both of which are a consequence of Hawking-radiation, to fix the value of the amplitude. The second makes no assumption pertaining to Hawking-radiation since this has not been verified experimentally to date; the amplitude is fixed by assuming the maximum possible value on $f_{\mathrm{PBH}} (=1)$. 

Additionally, for the narrow and broad spectra, we estimate upper limits on $f_{\mathrm{PBH}}$ assuming non-detection (where detection is assumed when the SNR exceeds a fiducial value of $2$). We find that we should be able to place non-trivial constraints from the second observing run, that don't depend on Hawking radiation. These become several orders of magnitude stronger than existing constraints that assume Hawking radiation by the end of the third observing run, and even stronger for design sensitivity of Advanced LIGO.

The article is organized as follows: We summarize how the expected SNR is calculated from the detector sensitivity curve and the energy-density spectrum $\Omega_{\mathrm{GW}}$ in Section \ref{sec:stochastic}. We then briefly describe the formation of stochastic GW background induced by primordial scalar perturbations in Section~\ref{sec:perturbations}.
 Section~\ref{sec:results} presents prospective SNR values for various detector sensitivities, and a range of ultralight PBHs associated with GWs in second order perturbation theory. Section~\ref{sec:conclusion} concludes the paper by summarizing and discussing the results, while also advocating a search for these PBHs in O1, O2 and O3 data (the latter could become available soon).

\section{The Stochastic Background and its detectability}\label{sec:stochastic}

Apart from the individually resolvable signals, like the ones being detected by LIGO and Virgo, we also expect a stochastic GW background to be present. This could be produced either by energetic processes in the early Universe, or by the incoherent superposition of many independent astrophysical signals whose amplitudes may be too weak to be detectable as individual sources (see, e.g., \cite{RomanoCornish2017} for a review). The stochastic GW background can be characterized completely from its statistical properties. The central quantity in the detection of this stochastic background is the spectrum of the GW energy density fraction, which is the fraction of the critical energy density $\rho_c$ required for a flat universe, as GWs, per logarithmic frequency bin \cite{Allen1997, AllenRomano1999}:
\begin{equation}
\Omega_{\mathrm{GW}}(f) = \frac{1}{\rho_c}\frac{d\rho_{\mathrm{GW}}}{d\log f},
\end{equation}
where $\rho_{\mathrm{GW}}$ is the energy density of GWs. Below, we show how $\Omega_{\mathrm{GW}}(f)$ is related to the GW polarizations. 

The standard TT gauge GW metric perturbation can be Fourier-expanded as a superposition of plane waves with frequency $f$ and propagation direction $\hat{n}$ \cite{ThraneRomano2013}:
\begin{eqnarray}
h_{ab}(t, \vec{x}) & = & \int_{-\infty}^{+\infty}df \int_{S^2}d^2\Omega_{\hat{n}} \sum_A h_A(f, \hat{n}) \, \mathrm{e}^{A}_{ab}(\hat{n}) \nonumber \\ 
 &\times& \exp[i \, 2\pi f \, (t - \hat{n}\cdot\vec{x}/c)],
\end{eqnarray}
where $d \Omega_n$ is the solid angle element, $A = {+, \times}$ denote the two polarizations of GWs, and $\mathrm{e}_{ab}^A $ denote the corresponding polarization tensors. 
Assuming that the stochastic GW background can be modelled as a zero-mean Gaussian random process, the mean $\langle h_A(f, \hat{n}) \rangle = 0$ and (co)variances $\langle h_A(f, \hat{n}) ~ h^*_{A'}(f, \hat{n}') \rangle$ of the random Fourier amplitudes become the defining characteristics of this background. The (co)variances are related to the energy density fraction via the GW power spectral density (PSD) $S_h(f)$ \cite{ThraneRomano2013}:
\begin{equation}
\langle h_A(f, \hat{n}) ~ h^{*}_{A'}(f, \hat{n}' \rangle = \frac{1}{16\pi} \, \delta(f - f') \, \delta_{AA'} \, \delta^2(\hat{n}, \hat{n}') \, S_h(f)
\end{equation}
with
\begin{equation}
S_h(f) = \frac{3H_0^2}{2\pi^2}\frac{\Omega_{\mathrm{GW}}(f)}{f^3}
\end{equation}

The detectors' response to a stochastic background is characterized by the overlap reduction function $\Gamma_{IJ}$, which acts as a transfer function between the GW PSD and the detector cross-power $C_{IJ}$, where $I, J$ are labels for two detectors. If $h(t)$ is the response strain of a detector to a GW metric perturbation $h_{ab}(t, \vec{x})$, then the detector cross power is related to the cross-correlation of the detector response strains as:
\begin{equation}
\langle \tilde{h}_I(f) ~ \tilde{h}_J^{*}(f') \rangle = \frac{1}{2}\delta(f - f') \, C_{IJ}(f)
\end{equation}
where $C_{IJ} = \Gamma_{IJ} \, S_h(f)$. The expected SNR can be evaluated from the GW PSD $S_h(f)$, the overlap reduction function $\Gamma_{IJ}$, as well as the detector noise PSDs $S_0(f)$ \cite{Allen1997, AllenRomano1999, ThraneRomano2013}:
\begin{equation}
\rho = \sqrt{2T\int_{f_\mathrm{low}}^{f_\mathrm{high}}df\left[\sum_{I=1}^{M}\sum_{J > I}^M \frac{\Gamma_{IJ}^2 \, S_h^2(f)}{S_{0I}(f) \, S_{0J}(f)}\right]}
\end{equation}
assuming all detectors have the same coincident observation time $T$. Above, $f_\mathrm{low} - f_\mathrm{high}$ denotes the sensitive frequency band of the detectors. In this work, we consider the stochastic GW background to be detectable if the SNR is greater than a fiducial threshold of 2.

\section{Stochastic GW background induced by primordial scalar pertubations}\label{sec:perturbations}
Here we briefly outline the generation of the stochastic GW background induced by scalar perturbations. Details of this derivation may be found in \cite{KohriTerada2018}, and a more concise version in \cite{Wangetal2019}. We assume that the GWs are produced in the radiation dominated era, and that the perturbations follow Gaussian statistics. We neglect any modifications to the stochastic background due to propagation through the intervening medium \cite{Domckeetal2020}. Furthermore, we also neglect potential enhancements to the background that could be produced during the matter dominated era \cite{KohriTerada2018}. These imply that the SNRs estimated from the energy density fraction derived here would be conservative.  

We have strong constraints on the curvature power spectrum, $P_{\zeta}(k)$, at scales $k \sim 10^{-4} - 10 \, {\rm Mpc}^{-1}$ coming from the cosmic microwave background (CMB) and the large scale structures, though the spectrum is almost unconstrained over smaller and larger scales. 
The exact shape of the curvature power spectrum over these unconstrained scales is currently unknown. Similarly, we practically have no constraints on the mass distribution of PBHs over the mass range corresponding to these scales. 
It is therefore natural to start with simplified models. Among the most commonly considered models are the monochromatic mass distribution of PBHs  though, it is generally agreed upon that such a distribution is unrealistic. 
However, there are models of inflation that can give rise to a narrow peak in the smaller-scale end in the curvature power spectrum that can, in turn, give rise to a nearly monochromatic PBH mass distribution. 
On the other hand, PBHs are possible with an extended mass distribution also, although one would need to provide its hyperparameters (such as, mean and variance) which are not known \emph{a priori}. A search for the induced stochastic GW background could inform and constrain these parameters.

We start by relating the wavenumber scale $k$ with the mass scale within the Hubble horizon, $M_{H}$
\begin{equation}
k \propto \frac{1}{\sqrt{M_H}}
\Big[g_{*\rho}(T(M_H)) \Big]^{1/4} \, \Big[g_{*s}(T(M_H)) \Big]^{-1/3},
\label{eq:mass-k_rel}
\end{equation}
where $T$ is the cosmic temperature, which is related to the mass scale as:
\begin{equation}
M_H = 12\left(\frac{10}{g_{*\rho}(T)}\right)^{1/2}\frac{M^3_G}{T^2}.
\end{equation}
Above, $M_G$ is the Plank mass, and $g_{*\rho}$,  $g_{*s}$ are the effective degrees of freedom of relativistic particles, which can be calculated assuming the Standard Model, as done in \cite{Saikawa2018} and used in \cite{KohriTerada2018, Wangetal2019}.

The GW energy density fraction as a function of the wavenumber can computed (semi-analytically) as follows \cite{KohriTerada2018, Wangetal2019}:
%
\begin{equation}
\Omega_{\mathrm{GW}}(k)\mid_{\mathrm{today}} = \frac{\Omega_{r0}}{24}\left(\frac{g_{*\rho}(T)}{g_{*\rho}(T_{eq})}\right)\left(\frac{g_{*s}(T)}{g_{*s}(T_{eq})}\right)^{-4/3}\left(\frac{k}{aH}\right)^2\overline{P}_h(\tau, k)
\label{eq:Omega_gw_k}
\end{equation}
%
where $a$ and $H$ denote, as usual, the scale factor and the Hubble parameter (which need to be evaluated at horizon entry), $T_{\mathrm{eq}}$ is the temperature of the Universe at the matter-radiation-equality epoch, and $\tau$ is the conformal time. The time averaged power spectrum $\overline{P_{h}}$ of the induced GWs is computed from the assumed curvature power spectrum (see, for example, equations D2 and D3 in \cite{Wangetal2019})

%
\begin{figure}[tb]
		\includegraphics[width=1\linewidth]{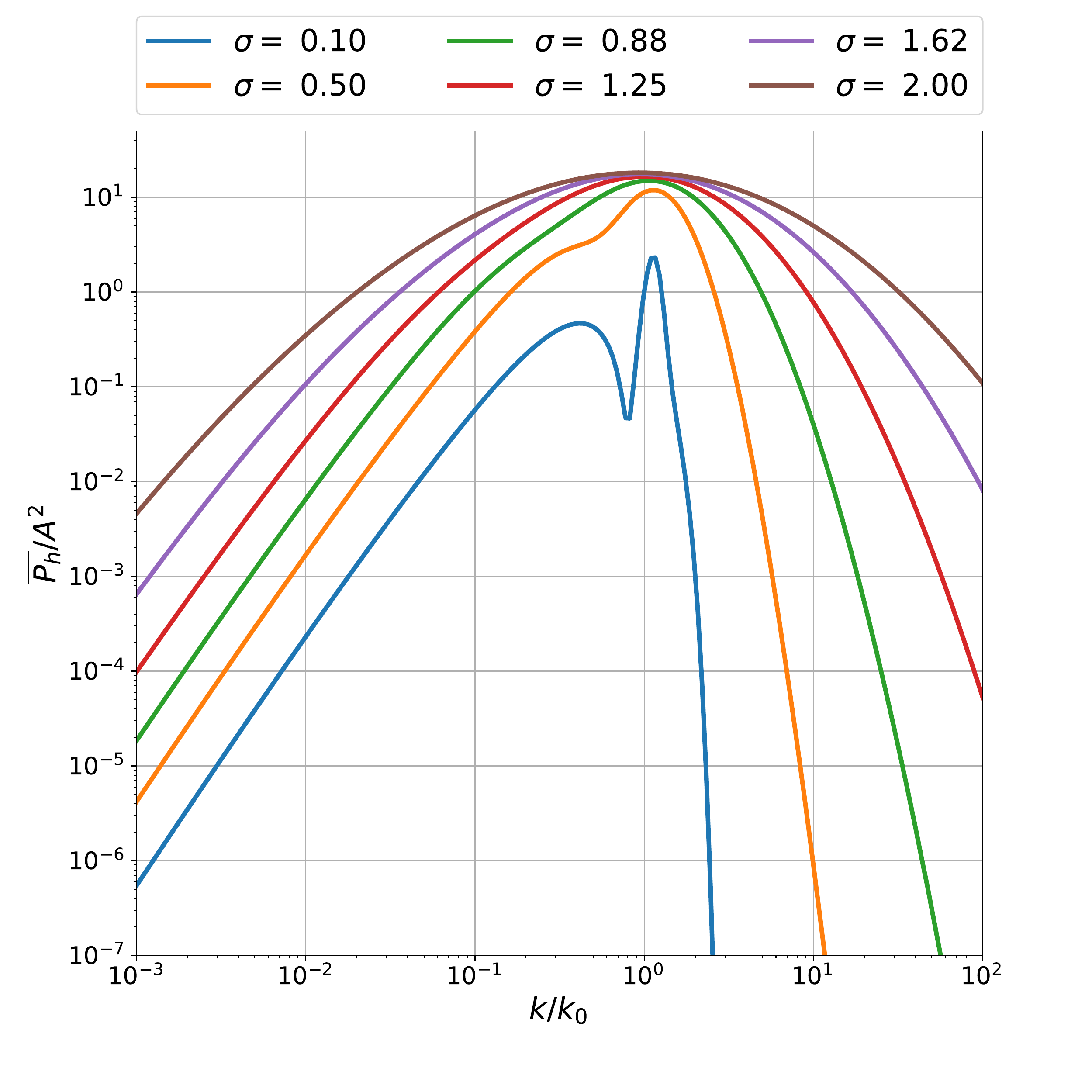}
		\caption{The time-averaged perturbation spectrum $\overline{P_h}$, rescaled by the amplitude $A$, as a function of the rescaled wavenumber $k/k_0$, for normal-in-log curvature power spectra, $P_{\zeta}(k)$. We vary the variance $\sigma$, and find that for sufficiently large values, the sharp feature present in the narrow $\overline{P_h}$ washes away.}
		\label{fig:Ph}
\end{figure}

\subsection{A ``Gaussian in $\log(k)$'' source}
\begin{figure}[tbh]
		\includegraphics[width=1\linewidth]{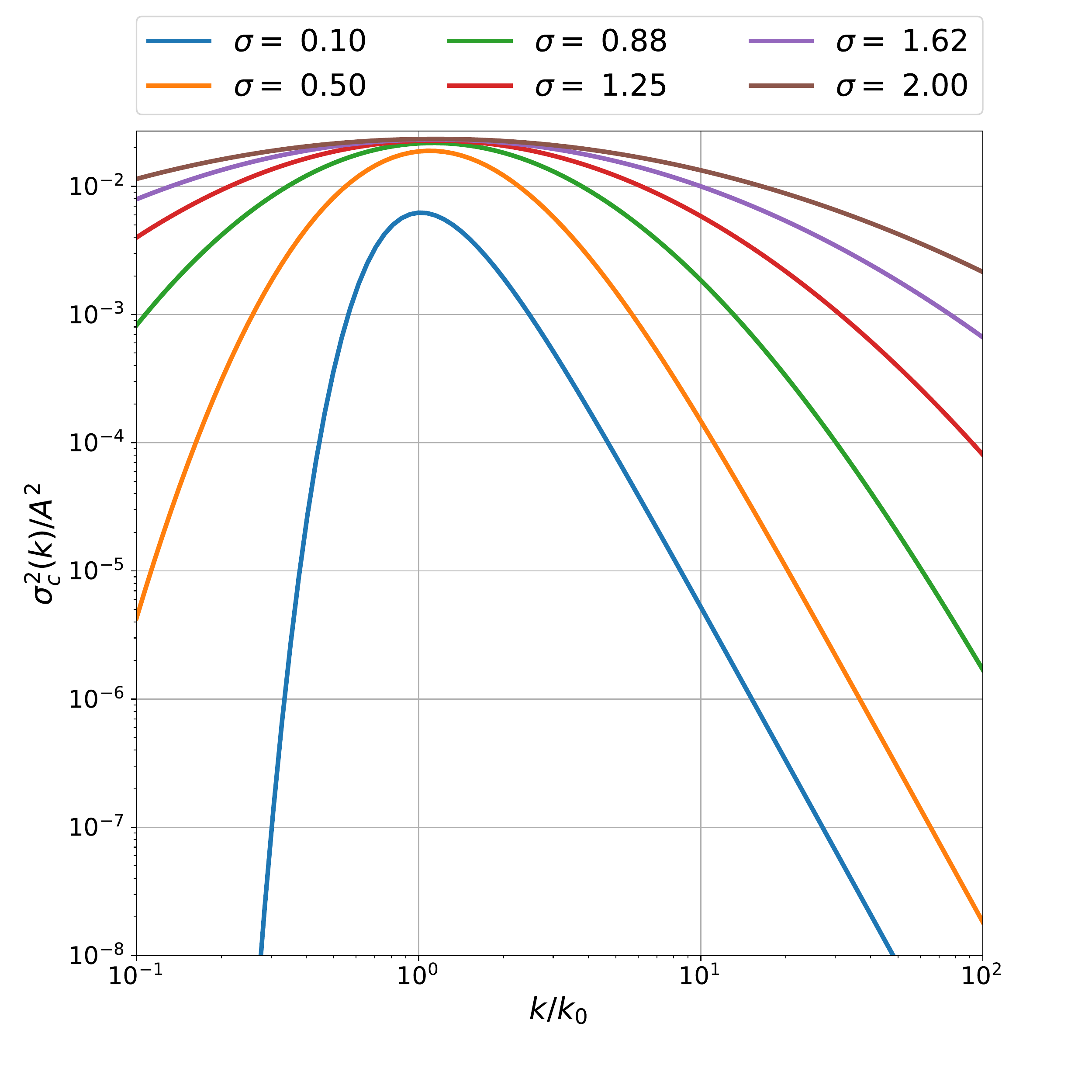}
		\caption{The coarse-grained perturbation $\sigma_c$ (the variance of the distribution of density fluctuations $\delta$) in the radiation dominated universe, rescaled by the amplitude $A$, as a function of the rescaled wavenumber $k/k_0$. Increasing the variance $\sigma$ of the normal-in-log curvature power spectrum increases the range of wavenumbers over which the $\sigma_c$ spans.}
		\label{fig:sigmac}
\end{figure}

An often considered, realistic choice for the curvature power spectrum $P_{\zeta}$ would be a normal distribution in $\log(k)$, centered on $\log(k_0)$, with a variance $\sigma^2$:
\begin{equation}
P_{\zeta}(k) = A\exp\left(-\frac{\left(\log(k/k_0)\right)^2}{2\sigma^2}\right)
\end{equation} 
The resulting $\Omega^{\mathrm{gauss}}_{\mathrm{GW}}$ needs to be evaluated numerically, by choosing values for $A, k_0$ and $\sigma$. In Fig.~\ref{fig:Ph}, we plot the corresponding $\overline{P_h}$, rescaled by $A^2$, for various values of the variance $\sigma^2$. We find that the sharp feature present for narrower Gaussians wash away with increasing values of $\sigma$. It is not clear a-priori which value of $\sigma$ should be used (see, for example, \cite{Carrilho2019} for lower limits on $\sigma$ in the context of single-field inflation theories, although values could be smaller in multi-field theories). We therefore select representative values; $\sigma = 0.88$ for ``broad'' distributions, and $\sigma = 0.1$ for ``narrow'' distributions.

\subsection{Fixing the amplitude $A$}
To fix the amplitude $A$, we follow Wang et al's \cite{Wangetal2019} prescription --- the mass-distribution $f(M)$ evaluated from the curvature power spectrum $P_{\zeta}$ is normalized to existing constraints (upper limits) on the fraction of dark matter $f_{\mathrm{PBH}}$ in the form of PBHs. The mass function is defined as follows:
\begin{equation}
f(M) \equiv \frac{1}{\Omega_{\mathrm{CDM}}}\frac{d\Omega_{\mathrm{PBH}}}{d\log M/M_{\odot}}
\end{equation}
where $\Omega_{\mathrm{CDM}}$ and $\Omega_\mathrm{PBH}$ are the cold dark matter and PBH energy densities. 
Computation of the PBH mass function from the sourcing 
curvature perturbation is non-trivial and a few formulations
have been proposed in the literature (recent progress along this 
direction is given in \cite{Suyama:2019npc, Germani:2019zez}).
Because of the strong exponential dependence of the PBH abundance
on $A$, $A$ would not be sensitive to the choice of the formulation
of the PBH mass function.
To be definite, in this paper we will use the Press-Schechter formalism \cite{PressSchechter1974} to compute the PBH mass function.

We briefly outline the method that can be used to acquire $f(M)$, given a density spectrum $P_{\zeta}(k)$. Assuming that PBHs are formed via critical collapse, the mass $M$ of a PBH is related to the Horizon mass scale $M_H$ and the amplitude of density fluctuation $\delta$ as follows:
\begin{equation}
M = KM_H\left(\delta -\delta_c \right)^{\gamma}
\end{equation}
where $K=3.3, \delta_c=0.19$ and $\gamma=0.36$ are numerical constants \cite{Wangetal2019, Young2020}. We assume a zero-mean Gaussian distribution of $\delta(M)$ with variance $\sigma^2_c(k(M_H))$ at a given horizon scale corresponding to $M_H$:
\begin{equation}
\mathcal{P}_{M_H}(\delta(M)) = \frac{1}{\sqrt{2\pi\sigma_c^2(k(M_H))}}\exp\left(-\frac{\delta^2(M)}{2\sigma^2_c(k(M_H))}\right)
\end{equation}
where $\sigma^2_c(k)$ is the variance of $\delta(M)$, and can be written in terms of curvature power spectrum $P_{\zeta}(k)$ as:
\begin{equation}
\sigma^2_c(k) = \frac{16}{81}\int_{-\infty}^{+\infty}\frac{q^4}{k^4} \mathcal{T}^2(q, 1/k) P_{\zeta}(q)d\log q w^2(q/k)
\end{equation}
where $w(q/k) = \exp(-q^2/(k^2))$ is a Gaussian window function \footnote{A few window functions have been considered in the literature; however, depending on the choice of window function, the corresponding value of $\delta_c$ needs to change so as to compensate for the resulting change in  $\sigma^2_c$. See \cite{Young2020} for details, including different choices of window functions and their corresponding $\delta_c$s.}, and $\mathcal{T}(q, \tau = 1/k) = 3(\sin y - y\cos y)/y^3, ~ y \equiv q\tau/\sqrt{3}$ is a transfer function. We plot $\sigma_c(k)$ in Fig.\ref{fig:sigmac} for a range of Gaussian power spectra, as a function of $k/k_0$ \footnote{The peak of $\sigma_c/A$ differs from $k/k_0 = 1$ by an amount that depends on the shape of $P_{\zeta}$. As a result, the PBH mass-scale must accordingly depend on $k$ rescaled by this shift in the peak. For a more detailed explanation, see \cite{Wangetal2019}.} and therefore the mass-function $f$. For the latter, we vary also the variance $\sigma^2$, and find that $\sigma_c(k)$ spans a wider range of $k/k_0$, as expected.

Using the Press-Schechter \cite{PressSchechter1974} formalism, the probability of PBH production, $\beta_{M_H}$ can be computed from the distribution on $\delta(M)$ as:
\begin{eqnarray}
\beta_{M_H} &=& \int_{\delta_c}^{\infty}\frac{M}{M_H}\mathcal{P}_{M_H}(\delta(M))d\delta(M) \\ \nonumber \\
&\equiv& \int_{-\infty}^{\infty}\tilde{\beta}_{M_H}(M)d\log M
\end{eqnarray}
For the assumed distribution $\mathcal{P}_{M_H}(\delta)$,  $\tilde{\beta}(M)$ is given by:
\begin{eqnarray}
\tilde{\beta}_{M_H}(M) & = & \frac{K}{\sqrt{2\pi\gamma^2 \sigma_c^2(k(M_H))}}
\left(\frac{M}{KM_H}\right)^{1 + 1/\gamma} \nonumber \\
 & \times & \exp\left(-\frac{1}{2\sigma_c^2(k(M_H))}\left(\delta_c + \left(\frac{M}{KM_H}\right)^{1/\gamma}\right)^2\right)
\end{eqnarray}
where $\tilde{\beta}_{M_H}$ is the distribution in the $\log$ of the PBH masses post critical collapse. The mass-function can now be computed from this distribution \footnote{It has been point out in the literature (see, e.g,  \cite{Kawasaki_2019, Luca_2019, Young2020} ) that a non linear relation between the comoving curvature perturbation and the density contrast leads to a correction of the amplitude of the power spectrum of a factor O(2) in order to give the same mass function. We incorporate this correction by multiplying our amplitudes by a factor of 2.}:
\begin{eqnarray}
f(M) & = & \frac{\Omega_m}{\Omega_{\mathrm{CDM}}}\int_{-\infty}^{\infty}\left(\frac{g_{*,\rho}(T(M_H))}{g_{*,\rho}(T_{eq})}\frac{g_{*,s}(T_{eq})}{g_{*,s}(T(M_H))}\frac{T(M_H)}{T_{eq}}\right) \nonumber \\ 
& \times & \tilde{\beta}_{M_H}(M)d\log M_H.
\end{eqnarray}
 
\begin{figure}[tb]
		\vspace{0.1cm}
		\includegraphics[width=1\linewidth]{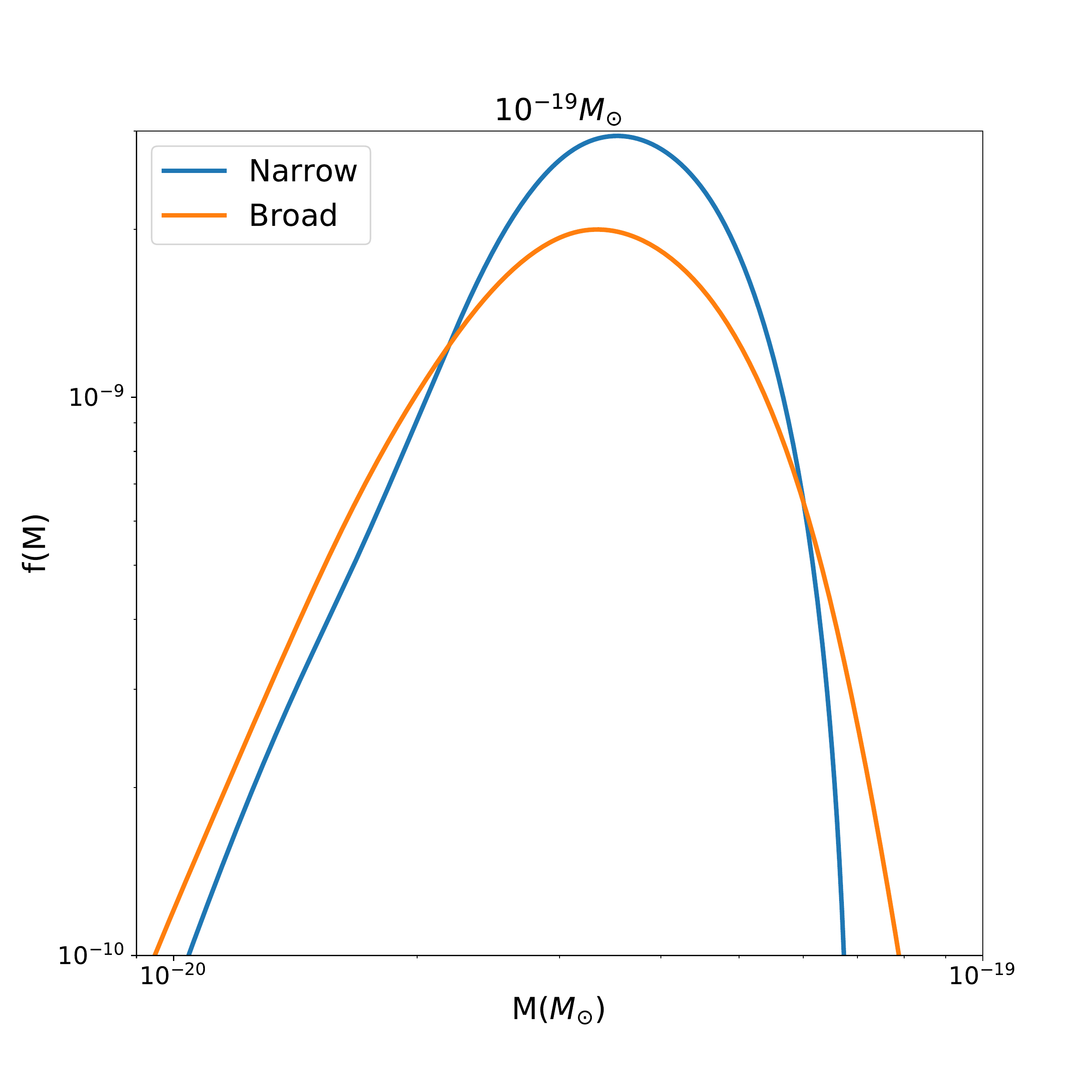}
		\caption{The mass function $f(M)$ for a central PBH mass of $10^{-19} M_{\odot}$, for the ``narrow'' ($\sigma = 0.1)$ and ``broad'' ($\sigma = 0.88$) Gaussian power spectra, $P_{\zeta}$. The power spectrum is normalized to the upper limit on $f_{\mathrm{PBH}}$ from other experiments (BBN, extra-galactic photon background, etc) that assume Hawking radiation (see for example \cite{Carretal2010} and references therein).}
		\label{fig:f_of_M}
\end{figure}

On the other hand $f_{\mathrm{PBH}}$ is given by:
\begin{equation}
f_{\mathrm{PBH}} = \int_{-\infty}^{\infty}f(M)d\log(M/M_{\odot}).
\end{equation}
For the mass-scales that we probe in this article, the upper limits on $f_\mathrm{PBH}$ have been placed, assuming Hawking radiation, from the extra-galactic photon background, as well as BBN. 
As the existing constraints, we will use the ones
given in \cite{2010PhRvD..81j4019C} which we show as
a black curve in Fig.~\ref{fig:fpbhUL}.
This curve is valid only for the PBHs with monochromatic mass function. A method to translate this curve to the upper limit
on the amplitude of an extended mass function was provided in \cite{Carr:2017jsz}, which we use in this paper to estimate maximal value for $A$. 
For illustrative purposes, in Fig.~\ref{fig:f_of_M}, we plot $f(M)$, for both the cases, the ``narrow'' ($\sigma = 0.1$) and ``broad'' ($\sigma = 0.88$) Gaussian peaks in the curvature power spectrum, for a mass-scale $10^{-19} M_{\odot}$. Conversely, since Hawking radiation has to date not been verified experimentally, we can also put an estimate on $A$ by setting $f_{\mathrm{PBH}} = 1$. The corresponding $\Omega_{\mathrm{GW}}$ would be the largest allowed in this formalism, and its non-detection would put Hawking-radiation-independent constraints on $f_{\mathrm{PBH}}$, the first of its kind in the narrow mass-range of ultralight PBHs probed in this paper.

Additionally, for the broad power spectrum, we estimate the upper limits on $f_{\mathrm{PBH}}$ assuming non-detection in all current and future runs of LIGO, with sensitivities pertaining to O1, O2, O3 and Design (see Fig.~\ref{fig:fpbhUL}). By the end of the third observing run, we find that the constraints on $f_{\mathrm{PBH}}$ improve by many orders of magnitude from those placed using Hawking radiation. Even by the second observing run, we are able to place non-trivial, Hawking-radiation-independent constraints, albeit weaker than those based on Hawking radiation.

\section{Results}\label{sec:results}
\begin{figure*}
\includegraphics[width=0.48\linewidth]{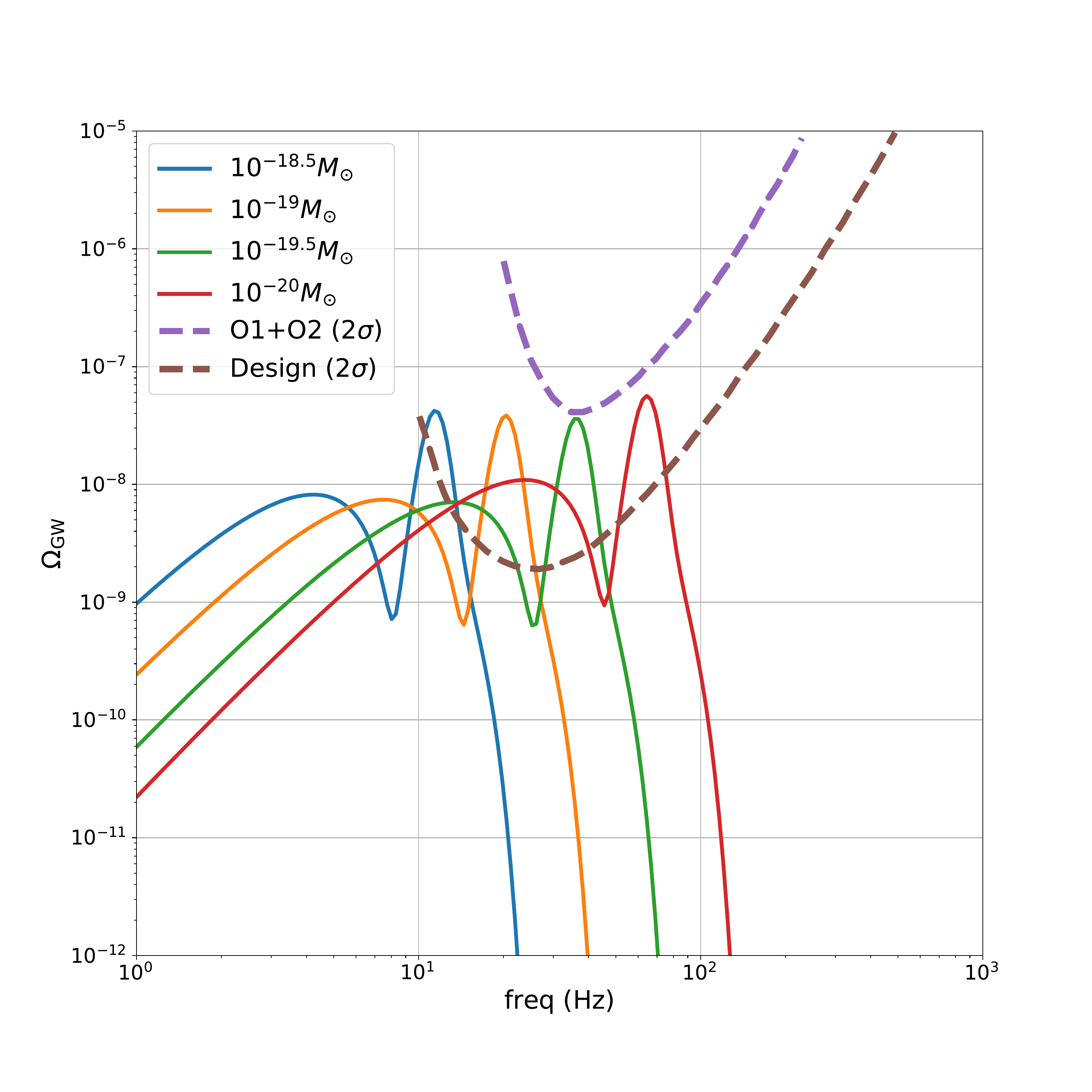}
\includegraphics[width=0.48\linewidth]{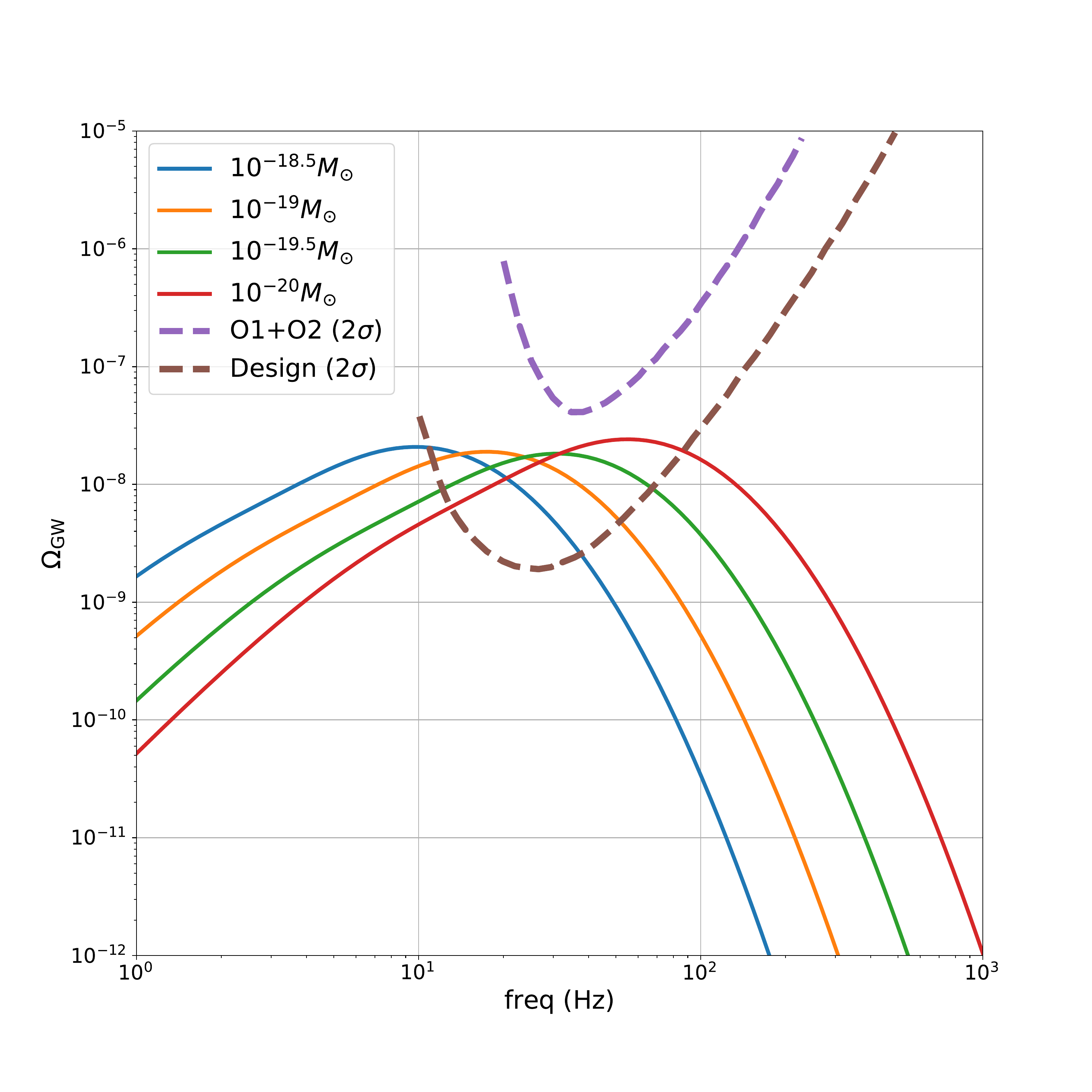}
\caption{The energy density fraction for the ``narrow'' ($\sigma = 0.1$) Gaussian peak in the curvature power spectrum case (left panel), and for the ``broad'' ($\sigma = 0.88$) Gaussian peak in the curvature power spectrum (right panel). We also plot the power-law-integrated curves \cite{ThraneRomano2013} for O1+O2 sensitivity, and design sensitivity of LIGO, taken from \cite{O1Stochastic}. Strictly speaking, these sensitivity curves are not valid for the energy density spectra considered here, especially the narrow case, since it cannot be modelled as a power law in the frequency. Nevertheless, these curves are indicative of the frequency range over which one might expect to acquire significant SNRs. Note that the power-law integrated curves correspond to detectability at $95\%$ confidence ($2\sigma$). The $\sigma$ in the legends of the plots should not be confused with the $\sigma$ associated with the Gaussian PBH mass distribution.}
\label{Omega_gw}
\end{figure*}

We evaluate expected SNRs for various sensitivities of the LIGO detectors, for choices of masses: $10^{-19}, 10^{-19.5}, 10^{-19.5}, 10^{-20} M_{\odot}$, and set an SNR threshold of 2 for detectability. As shown in Fig.~\ref{Omega_gw}, these masses are associated with $\Omega_{\mathrm{GW}}$s that fall within LIGO's sensitivity band, both for the ``narrow'' and ``broad'' Gaussian peaks in the curvature power spectrum. For the case of the ``broad'' Gaussian peak, we take $\sigma=0.88$ as a representative value given that $\sigma={\cal O}(1)$ is a natural possibility. On the other hand, for the ``narrow'' peak, we take a representative value of $\sigma = 0.1$.
We also plot the power-law integrated sensitivity curves of LIGO, which represent sensitivity curves for a stochastic GW background whose $\Omega_{\mathrm{GW}}(f)$ can be approximated as a power-law in the frequency.  

We consider two cases, as described in an earlier section. For the first case, the amplitude of the curvature power spectrum, $A$, is fixed so as to respect the upper limits on the mass-functions set by constraints on $f_{\mathrm{PBH}}$ by other experiments that assume Hawking radiation. For the second case, we allow the amplitude to take its maximum possible value, as determined by $f_{\mathrm{PBH}} = 1$. All the SNRs are tabulated in Table \ref{tab:expected_snr}. We restrict ourselves to a two-detector network consisting of LIGO-Hanford and LIGO-Livingston. For the O1 and O2 observing runs, we use the publicly available PSDs estimated by the LIGO-Virgo collaboration \cite{GWOSC}. For O3 sensitivity, we use the PSD from \cite{O3Sensitivity}; for Advanced LIGO's \cite{aLIGO} Design sensitivity, we use the projected PSD \cite{DesignSensitivity}.

\begin{table}
\caption{Expected SNRs of the stochastic GW background corresponding to various PBH masses (indicated in the first column) and two assumed shapes (``Narrow'': $\sigma = 0.1$, ``Broad'': $\sigma = 0.88$) of the primordial power spectrum (second column) in various observing runs of LIGO (third column).
The lower limit on the frequency range is kept at $10$ Hz for O1, O2 PSDs, 20 Hz for O3 PSD, and $5$ Hz for the Design PSD. The amplitudes $A$ of the curvature power spectrum  set by using existing constraints on $f_{\mathrm{PBH}}$ are shown in the fourth column. Those set by using the maximum allowed value for amplitude $A$, when $f_{\mathrm{PBH}} = 1$, are tabulated parenthetically.} 
\centering 
\begin{tabular}{c c c c c} 
\hline\hline 
Mass ($M_{\odot}$) & $P_{\xi}(k)$ & Obs. Run & $A \times 10^{-2}$ & SNR\\ [0.5ex] 
\hline 
$10^{-18.5}$ & Narrow & O1+O2 & 10.9 (16.1) & 0.0 (0.0) \\ 
$10^{-18.5}$ & Narrow & O3 & 10.9 (16.1) & 0.0 (0.0) \\
$10^{-18.5}$ & Narrow & Design & 10.9 (16.1) & 5.7 (12.5) \\
\hline
$10^{-19}$ & Narrow & O1+O2 & 10.3 (15.9) & 0.3 (0.6) \\ 
$10^{-19}$ & Narrow & O3 & 10.3 (15.9) & 2.0 (4.6) \\
$10^{-19}$ & Narrow & Design & 10.3 (15.9) & 15.2 (35.8) \\
\hline
$10^{-19.5}$ & Narrow & O1+O2 & 10.1 (15.6) & 1.9 (4.5) \\ 
$10^{-19.5}$ & Narrow & O3 & 10.1 (15.6) & 4.3 (10.2) \\
$10^{-19.5}$ & Narrow & Design & 10.1 (15.6) & 15.2 (36.4) \\
\hline
$10^{-20}$ & Narrow & O1+O2 & 12.5 (15.4) & 0.7 (1.0) \\ 
$10^{-20}$ & Narrow & O3 & 12.5 (15.4) & 2.1 (3.1) \\
$10^{-20}$ & Narrow & Design & 12.5 (15.4) & 9.8 (14.8) \\[1ex]
\hline\hline
$10^{-18.5}$ & Broad & O1+O2 & 3.07 (4.52) & 0.4 (0.8) \\ 
$10^{-18.5}$ & Broad & O3 & 3.07 (4.52) & 1.3 (2.9) \\
$10^{-18.5}$ & Broad & Design & 3.07 (4.52) & 10.4 (22.5) \\
\hline
$10^{-19}$ & Broad & O1+O2 & 2.92 (4.45) & 1.0 (2.4) \\ 
$10^{-19}$ & Broad & O3 & 2.92 (4.45) &  3.2 (7.3) \\
$10^{-19}$ & Broad & Design & 2.92 (4.45) & 16.8 (38.9) \\
\hline
$10^{-19.5}$ & Broad & O1+O2 & 2.87 (4.38) & 1.6 (3.7) \\ 
$10^{-19.5}$ & Broad & O3 & 2.87 (4.38) & 4.3 (9.9) \\
$10^{-19.5}$ & Broad & Design & 2.87 (4.38) & 18.2 (42.4) \\
\hline
$10^{-20}$ & Broad & O1+O2 & 3.30 (4.32) & 1.9 (3.2) \\ 
$10^{-20}$ & Broad & O3 & 3.30 (4.32) & 4.6 (7.9) \\
$10^{-20}$ & Broad & Design & 3.30 (4.32) & 17.6 (30.1) \\[1ex]
\hline 
\end{tabular}
\label{tab:expected_snr} 
\end{table}

\begin{figure*}
\includegraphics[width=0.48\linewidth]{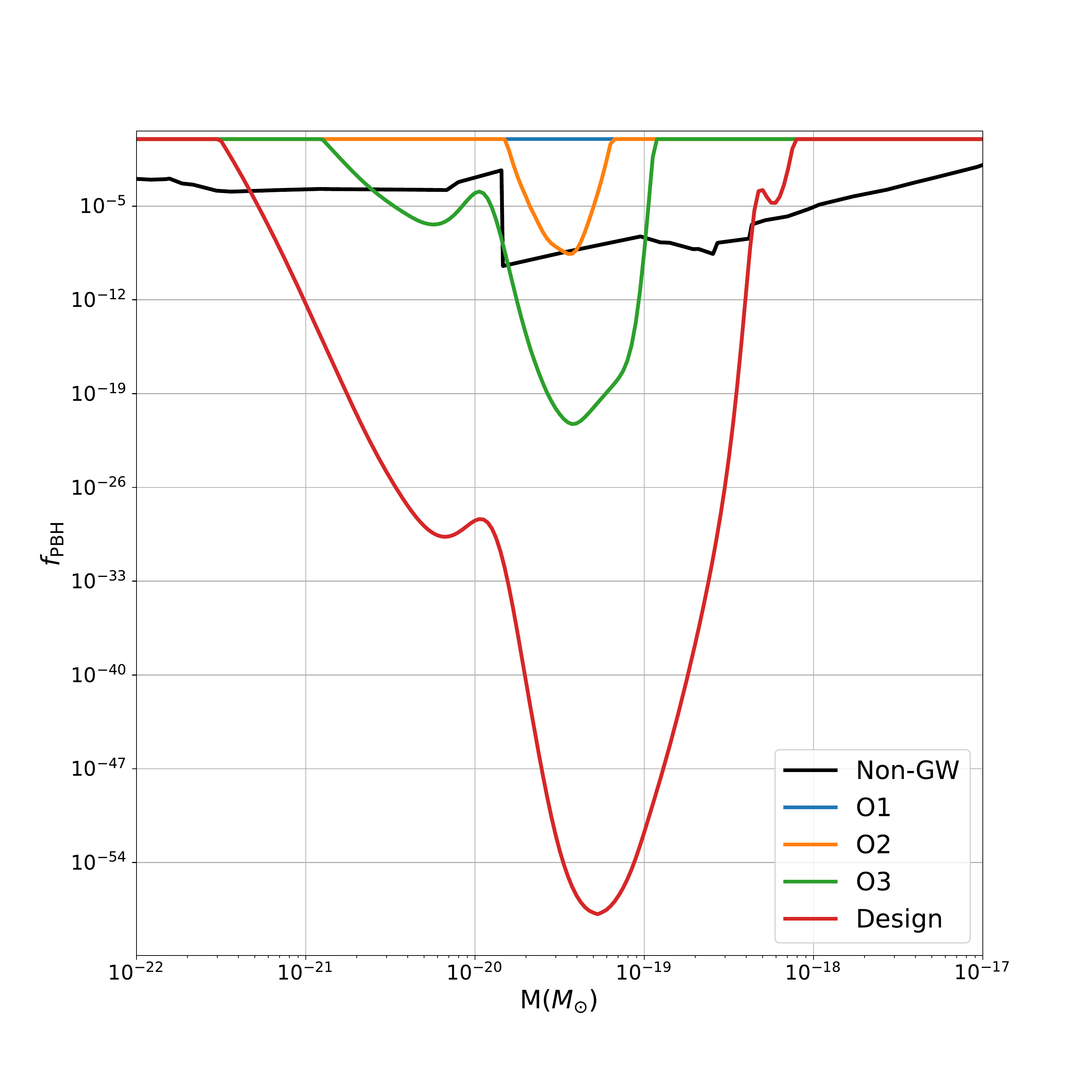}
\includegraphics[width=0.48\linewidth]{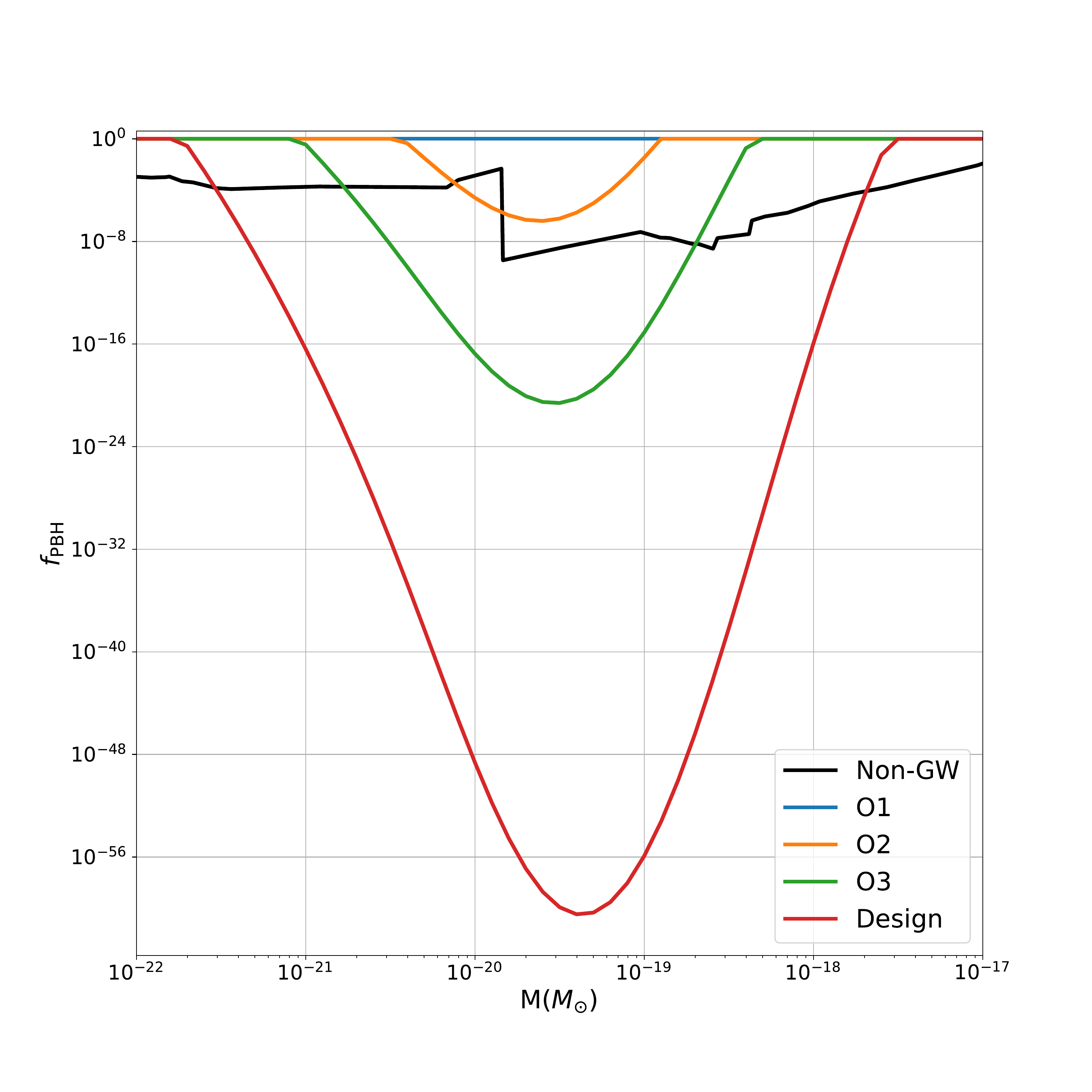}
\caption{Expected upper limits on $f_{\mathrm{PBH}}$ as a function of the PBH mass, assuming the ``narrow'' (left panel) and ``broad'' (right panel) Gaussian power spectra ($\sigma=0.1$ and $\sigma = 0.88$, respectively), and non-detection. From O2, we already expect to get the first ever Hawking-radiation independent constraints, albeit weaker than existing constraints that assume Hawking radiation (denoted as ``Non-GW'' above; see, e.g, \cite{Carretal2010, Carr:2017jsz}). From O3 and design-sensitivity runs, these constraints are expected to improve significantly, becoming many orders of magnitude stronger than existing constraints. The drastic change in constraints between observing runs can be attributed to the sensitive dependence of the mass-function $f(M)$ on the amplitude $A$ of the curvature power spectrum.}
\label{fig:fpbhUL}
\end{figure*}

We find that the detectability of the GW background depends not only on the sensitivity of the detectors, but also on the choice of the form of curvature power spectrum, as well as its amplitude $A$. Among the PBH masses considered, $10^{-19, -19.5} M_{\odot} $ correspond to the most optimistic scenarios in terms of the expected SNRs. With the O1+O2 sensitivity, the expected SNRs cross the threshold of 2 for both the narrow and broad spectra, only if the amplitude is allowed to takes its maximum possible value ($f_{\mathrm{PBH}} = 1$). On the other hand, for the O3 sensitivity, expected SNRs are above the threshold for both types of spectra, and a wider PBH mass range, even for a less optimistic choice of $A$ (its maximum possible value that is not ruled out by current experiments). By the time the LIGO detectors reach design sensitivity, the background corresponding to all PBH masses considered here should be detectable with significant SNRs, if the curvature power spectrum has the amplitude $A$ that we assume. Conversely, a non-detection would put strong constraints on the amplitude $A$ of the curvature power spectrum as well as the abundances of the corresponding PBHs, independent of Hawking radiation considerations.
To demonstrate this, we plot in Fig. \ref{fig:fpbhUL} the upper limits on $f_{\mathrm{PBH}}$ assuming non-detection, for both the narrow and broad Gaussian power spectra and for all observing runs. We find that by the third observing run, we get constraints that are several orders of magnitude stronger than existing constraints. 

It is worth mentioning that all the would-be constraints we obtained are for the curvature perturbations obeying Gaussian statistics. Non-Gaussian curvature perturbations will tighten or loosen those constraints depending on the character of non-Gaussianity. For instance, the curvature perturbations obeying the local-type non-Gaussianity parametrized by the $f_{NL}$ parameter, a mildly negative (positive) $f_{NL}$ weakens (tightens) the upper limit on $A$ \cite{Young:2013oia}.
If the non-Gaussianity is incorporated, the observational data will provide the joint constraints on $A$ and the non-Gaussianity parameter.

\section{Summary and Outlook}\label{sec:conclusion}
In this work, we explore the possibility of detecting the stochastic GW background produced by primordial curvature perturbations, which are associated with the formation of PBHs. These GWs, produced by scalar-tensor mode coupling in the second order perturbation theory, can be modelled assuming a shape for the curvature power spectrum $P_{\zeta}(k)$. We consider two Gaussian models, in $\log(k)$, for the primordial curvature power spectrum, with mean $\log(k_0)$, and variance $\sigma^2$. The ``narrow'' Gaussian has $\sigma = 0.1$ and the ``broad'' Gaussian has $\sigma = 0.88$. Following the prescription of Wang et al~\cite{Wangetal2019}, we evaluate the energy density fraction $\Omega_\mathrm{GW}(f)$ from the power spectrum $P_{\zeta}$. We find that $\Omega_\mathrm{GW}(f)$ falls within the LIGO detectors' sensitivity bands for a narrow range of ultralight PBH masses $\lesssim 10^{15}$g. Motivated by this, we compute the expected SNRs, for both cases, assuming $\Omega_{\mathrm{GW}}$s associated with PBH masses $10^{-18.5}, 10^{-19}, 10^{-19.5}$ and $10^{-20} M_{\odot}$. We consider sensitivities associated with LIGO observing runs O1+O2, O3 (projected) as well as the design sensitivity (projected). 

The power-spectra have a free amplitude parameter $A$, which we fix in two ways. The first normalizes the mass-function $f(M)$ to the upper limits on $f_{\mathrm{PBH}}$ from other experiments that assume Hawking radiation, in the ultralight PBH regime. The second allows $f_\mathrm{PBH}$ to attain its maximum possible value of unity, not relying on Hawking-radiation based constraints since it has thus far not been verified experimentally.

We then compute the expected SNRs, which we tabulate in Table \ref{tab:expected_snr}. Among the four masses considered, the most optimistic from a detection perspective are $10^{-19, -19.5} M_{\odot}$, which produce a significant expected SNR even for O1+O2 sensitivity. However, it must noted that this is only if Hawking radiation based constraints are not considered. At O3 and design sensitivity, all masses are considered detectable, for both choices of power spectra, and even after considering the current constraints on $f_{\mathrm{PBH}}$ (which limits the value of $A$). 

The expected SNRs advocate searching for Gaussian spectra for a broad range of widths, in all observing runs (both completed and upcoming), and, assuming non-detection, place upper limits on the amplitude $A$ of the primordial power spectrum as well as the abundance of primordial black holes in the relevant mass range. To illustrate this, we place constraints on $f_{\mathrm{PBH}}$ as a function of the PBH mass, for all observing runs, and both the narrow and broad Gaussian power spectra. We find that, with O2, we can already place non-trivial Hawking-radiation-independent constraints, albeit weaker than existing ones that rely on Hawking radiation. With O3, we get constraints that are several orders of magnitude stronger than the existing ones, which get even stronger with the design-sensitivity run. 

At design sensitivity, the stochastic background from stellar mass BBH mergers would also be detectable \cite{O1Stochastic}. Therefore, a significant SNR would not in itself suggest the detection of the induced stochastic background associated with PBHs. Nevertheless, the difference in the shape of $\Omega_{\mathrm{GW}}$ from each of these sources should enable us to distinguish the primordial background from the astrophysical background. 

The detectability of the induced stochastic GW background in current and future sensitivities of the LIGO detectors has been either touched upon in previous work (see \cite{Ballesterosetal2020}), or explored in more detail assuming a log-normal power spectrum \cite{InomataTomohiro2019}. The latter work focuses on the very important topic of constructing sensitivity (power-law integrated) curves \cite{ThraneRomano2013} for such stochastic backgrounds, for log-normal power spectra and LIGO at design sensitivity. They don't however explicitly compute SNRs for a range of current and future sensitivities of LIGO, nor evaluate values of the amplitude of the curvature power spectrum with and without constraints on $f_{\mathrm{PBH}}$ from other experiments. Additionally, our results incorporate more recent developments (for example, the amplification of the amplitude $A$ due to the non-linear connection between comoving curvature perturbation and the density contrast) in the relation between the curvature power spectrum, and the PBH abundance. We are currently in the process of searching for these Gaussian stochastic backgrounds in the O1+O2 data, and evaluating the constraints on $A$ and $f_{\mathrm{PBH}}$, which we hope to report soon.

\paragraph*{Acknowledgements:} 

We would like to thank Shivaraj Kandhasamy, as well as the anonymous referee of our manuscript, for their meticulous review and well considered comments. We also thank Sai Wang, Kazunori Kohri, Soichiro Morisaki, and
Joseph Romano for illuminating discussions. We would also like to acknowledge the Summer School on Gravitational Wave Astronomy (ICTS/gws2019/07) organized by the International
Centre for Theoretical Science (ICTS), TIFR, which served as the genesis for this project. SJK’s, KLP’s and PA’s research
was supported by the Department of Atomic Energy, Government of India. In addition, SJK’s research was supported by
the Simons Foundation through a Targeted Grant to ICTS. PA’s research was supported by the Max Planck Society through a Max Planck Partner Group at ICTS and by the Canadian
Institute for Advanced Research through the CIFAR Azrieli Global Scholars program. TS was supported by the MEXT Grant-in-Aid for Scientific Research on Innovative Areas No. 17H06359, No. 18H04338, and No. 19K03864. This research has made use of data, software and/or web tools obtained from the Gravitational Wave Open Science Center (https://www.gwopenscience.org), a service of LIGO Laboratory, the LIGO Scientific Collaboration and the Virgo Collaboration. LIGO is funded by the U.S. National Science Foundation. Virgo is funded by the French Centre National de Recherche Scientifique (CNRS), the Italian Istituto Nazionale della Fisica Nucleare (INFN) and the Dutch Nikhef, with contributions by Polish and Hungarian institutes.


%


\end{document}